\documentclass[floatfix,showpacs]{revtex4}
\usepackage{epsfig}
\usepackage{graphicx}
\usepackage[german, english]{babel}
\usepackage{amsmath}
\usepackage{amssymb}
\usepackage{ifthen}
\usepackage{epsfig}
\newcounter{fig}   \newcommand{\lbfig}[1]{\refstepcounter{fig}
\label{#1} } 
\newcommand{\vphi}{\varphi}

\begin{document}

\title{Monopole-Monopole solutions of Einstein-Yang-Mills-Higgs Theory} 

\author{Yasha Shnir} 

\affiliation{Institut f\"ur Physik, Universit\"at Oldenburg,
D-26111, Oldenburg, Germany}

\pacs{04.20.Jb,04.40Nr}

\begin{abstract}
New static regular axially symmetric solutions of  
$SU(2)$ Yang-Mills-Higgs theory are constructed. They are 
asymptotically flat and represent gravitating monopole-monopole 
pairs. The solutions form two branches linked to the second 
Bartnik-McKinnon solution on upper mass branch and to the  
monopole-monopole configuration in flat space on the lower branch, respectively. 
\end{abstract}


\maketitle


\section{Introduction}

The nontrivial vacuum structure of $SU(2)$ Yang-Mills-Higgs (YMH) theory
allows for the existence of regular non-perturbative finite energy 
solutions, such as spherically symmetric monopoles \cite{mono}, various 
axially symmetric multimonopoles \cite{WeinbergGuth,RebbiRossi,mmono,KKT} and   
monopole-antimonopole systems \cite{Rueber,mapKK,KKS}. In the  
Bogomol'nyi-Prasad-Sommerfield  (BPS) limit of vanishing scalar potential, 
axially symmetric multimonopole configurations 
are known analytically \cite{mmono}. In these solutions the nodes of the Higgs 
field are superimposed at a single point. Since in the BPS limit
the repulsive and attractive forces between monopoles exactly compensate, 
the BPS monopoles experience no net interaction \cite{Manton77}. Thus, 
the BPS configuration with multiple node at the origin can be continuously
deformed into a system of individual monopoles with unit topological charge 
(see, e.g., \cite{Manton-book,S-book}). However, as scalar field becomes massive, 
the fine balance of forces between the monopoles is broken and there is only   
repulsion between non-BPS multimonopole solutions \cite{KKT}. 
On the other hand, there are 
axially symmetric saddlepoint solutions of the YMH theory, 
where the Higgs field vanishes at several isolated points on the
symmetry axis \cite{Rueber,mapKK,KKS}. Simplest such a solution represent 
a monopole-antimonopole (MA) pair, a magnetic dipole \cite{Rueber,mapKK}.

When gravity is coupled to YMH theory, branches of gravitating solutions arise 
\cite{gmono,HKK,MAP}. The lower branch emerges from the flat space configurations
as coupling constant $\alpha$, entering the Einstein-Yang-Mills-Higgs (EYMH) equations,
is increases from zero. However, there is a difference between behavior of the 
gravitating solutions with a single or superimposed nodes of the scalar field \cite{gmono,HKK}
and gravitating monopole-antimonopole chains and vortex rings \cite{MAP}. While in the former 
case the lower branch ends at a critical value, beyond which the core of the configuration 
would be smaller than the Schwarzschild radius \cite{gmono,HKK}, for the gravitating 
monopole-antimonopole chains and vortex rings the second branch emerges which is extended 
back to  $\alpha =0$ \cite{MAP} . In this limit the configurations shrink to zero size and 
the scaled solutions approach corresponding solutions of pure  Einstein-Yang-Mills (EYM) theory 
\cite{BM,KK,IKKS}. 

It was mentioned that the additional attraction in the YMH system due to the 
coupling to gravity 
also allows for bound monopoles with superimposed zeros of the scalar field which  
are not present in flat space \cite{HKK}. However, one can conjecture if 
the EYMH theory allows for further 
static axially-symmetric solutions representing, for example,  
separated monopole-monopole (MM) pair. Evidently, beyond BPS limit these solutions do 
not possess counterparts 
in flat space but gravitational attraction may reinforce 
the effect of the scalar interaction and a bound state 
of the gravitating monopoles can exist. .   
 
In this letter we report the existence of one such new solution representing 
MM pair. On the 
upper branch it is related to the second Bartnik-McKinnon solution with two zeros \cite{BM}. 
The properties of the gravitating monopole pair are compared with those of 
MA pair, which on the 
upper branch is linked to the first  Bartnik-McKinnon solution with one zero, 
and with solution for the gravitating charge 2 axially symmetric monopole, 
linked to the extremal Reissner-Nordstr\"om 
black hole solution. 

In section II we present the Lagrangian of the EYMH theory, the axially symmetric ansatz 
and the boundary conditions. In section III we discuss the properties of the gravitating 
MM pair.

\section{\bf Einstein-Yang-Mills-Higgs model and axially symmetric ansatz}
We consider the  $SU(2)$ Einstein-Yang-Mills-Higgs 
theory with action 
\begin{equation} \label{action}
S =  \int \left\{ \frac{R}{16\pi G} 
-\frac{1}{2} {\rm Tr} 
\,\left( F_{\mu\nu}F^{\mu\nu} \right)
-\frac{1}{4} {\rm Tr}
\left(  D_\mu \Phi\, D^\mu \Phi  \right)
-\frac{\lambda}{8} 
 {\rm Tr} \left(\Phi^2 - \eta^2 \right)^2 
\right\}
\sqrt{- g} d^4 x
\end{equation}
Here $G$ and $\lambda$ denote the gravitational and scalar coupling constants,
respectively, $\eta$ is the vacuum expectation value of the Higgs field and $R$ is 
Ricci scalar. 

In isotropic coordinates the static axially symmetric metric reads \cite{KK,HKK,IKKS}
\begin{equation}
ds^2=
  - f dt^2 +  \frac{m}{f} d r^2 + \frac{m r^2}{f} d \theta^2
           +  \frac{l r^2 \sin^2 \theta}{f} d\vphi^2
\ , \label{metric2} \end{equation}
where the metric functions
$f$, $m$ and $l$ are functions of
the coordinates $r$ and $\theta$, only.
The $z$-axis ($\theta=0, \pi$) represents the symmetry axis.
Regularity on this axis requires $m=l$ there. 

For the gauge and the Higgs field we employ the known ansatz 
  \cite{KKS}
\begin{eqnarray}
A_\mu dx^\mu
& = & 
\left( \frac{K_1}{r} dr + (1-K_2)d\theta\right)\frac{\tau_\vphi^{(n)}}{2e}
- n \sin\theta \left( K_3\frac{\tau_r^{(n,m)}}{2e}
                     +(1-K_4)\frac{\tau_\theta^{(n,m)}}{2e}\right) d\vphi
\ , \label{ansatzA} \\
\Phi
& = &
\eta \left( \Phi_1\tau_r^{(n,m)}+ \Phi_2\tau_\theta^{(n,m)} \right) \  .
\label{ansatzPhi}
\end{eqnarray}
where the $su(2)$ matrices
$\tau_r^{(n,m)}$, $\tau_\theta^{(n,m)}$, and $\tau_\vphi^{(n)}$
are defined as products of the spatial unit vectors
\begin{eqnarray}
{\hat e}_r^{(n,m)} & = & \left(
\sin(m\theta) \cos(n\vphi), \sin(m\theta)\sin(n\vphi), \cos(m\theta)
\right)\ , \nonumber \\
{\hat e}_\theta^{(n,m)} & = & \left(
\cos(m\theta) \cos(n\vphi), \cos(m\theta)\sin(n\vphi), -\sin(m\theta)
\right)\ , \nonumber \\
{\hat e}_\vphi^{(n)} & = & \left( -\sin(n\vphi), \cos(n\vphi), 0 \right)\ ,
\label{unit_e}
\end{eqnarray}
with the Pauli matrices $\tau^a$. 

For $m=2$, $n=1$ the ansatz corresponds to the one for the 
MA pair solutions \cite{Rueber,mapKK,MAP},
while for $m=1$, $n>1$ it corresponds to
the ansatz for axially symmetric multimonopoles \cite{RebbiRossi,KKT,HKK}.
In particular, for $m=1$, $n=2$ we have non-BPS extension of the 
charge 2 monopole solution \cite{mmono}. 

The four gauge field functions $K_i$ and two Higgs field functions 
$\Phi_i$ depend on the coordinates $r$ and $\theta$, only.
To construct regular solutions we have to fix the gauge 
 condition 
$ r \partial_r K_1 - \partial_\theta K_2 = 0 $ \cite{KKT}.
Further, we introduce the dimensionless coordinate $x = er\eta$ and rescale the 
Higgs field as $\Phi  \to \Phi/\eta$. Then the dimensionless coupling $\alpha$, 
$\alpha^2 = 4\pi G \eta^2 $ enters the equations.

To obtain asymptotically flat solutions which are regular and corresponds to the 
gravitating MM pair, we need to impose the boundary conditions. 
Regularity of the solutions at the origin ($r=0$) 
requires for the metric functions the boundary conditions 
$
\partial_r f(r,\theta)|_{r=0}= 
\partial_r m(r,\theta)|_{r=0}= 
\partial_r l(r,\theta)|_{r=0}= 0 \ , 
$
whereas the gauge field functions $K_i$ satisfy
$
K_1(0,\theta)= K_3(0,\theta)= 0, \ 
K_2(0,\theta)= K_4(0,\theta)= 1 \  ,
$
and the Higgs field functions $\Phi_i$ satisfy
\begin{equation}
\sin \theta \ \Phi_1(0,\theta) + \cos \theta \ \Phi_2(0,\theta) = 0 \ ,
\end{equation}
\begin{equation}
\left.\partial_r\left[\cos \theta \ \Phi_1(r,\theta)
              - \sin \theta \ \Phi_2(r,\theta)\right] \right|_{r=0} = 0 \ .
\end{equation}
These conditions are the same both for MA-pair and MM-pair. For the charge 2 monopole 
both $\Phi_1$ and $\Phi_2$ must vanish at the origin.  

The boundary conditions at infinity shall provide correct   
asymptotic behavior of the  
fields depending on the configuration. Evidently, asymptotical flatness requires 
$ 
f \longrightarrow 1 , \  
m \longrightarrow 1 , \ 
l \longrightarrow 1 \ 
$
for any solution. Also the Higgs field at infinity have to approach the `hedgehog' 
shape, i.e., 
$\Phi_1\longrightarrow  1, \  \Phi_2 \longrightarrow 0 \ $. 
But the boundary conditions on the gauge functions can be different. 

To construct MA-pair, which is a deformation of the 
topologically trivial sector, 
the gauge field at infinity required to tend to a pure gauge 
$
A_\mu \ \longrightarrow  \ i \partial_\mu U U^\dagger \ ,
$
where  $U = \exp\{-i \theta\tau_\vphi^{(n)}\}$ \cite{Rueber,mapKK}. 
In terms of the functions $K_i$ these boundary
conditions read: 
\begin{equation}  \label{bound_MA}
K_1 \longrightarrow 0 \ , \ \ \ \
K_2 \longrightarrow - 1  \ , \ \ \ \
K_3 \longrightarrow 2 \sin \theta  \ , \ \ \ \
K_4 \longrightarrow 1- 2 \cos \theta \ , 
\end{equation}
whereas for the charge 2 monopole we required 
\begin{equation}  \label{bound_2M}
K_1 \longrightarrow 0 \ , \ \ \ \
K_2 \longrightarrow 0  \ , \ \ \ \
K_3 \longrightarrow 0  \ , \ \ \ \
K_4 \longrightarrow 0  \ .
\end{equation}

To construct monopole-monopole pair configuration on the same axially symmetric
ansatz (\ref{ansatzA}),(\ref{ansatzPhi}), let us note that the 
multimonopoles can be nicely described in terms of 
the effective electromagnetic quantities \cite{KKS,S05}, like magnetic charge 
\begin{equation}  \label{charge}
g = \frac{1}{4\pi }\int \frac{1}{2}
{\rm Tr} \, \left( F_{ij} D_k \Phi \right)\varepsilon_{ijk} d^3 r
\    \end{equation}
and the dimensionless 
magnetic dipole moment $\mu$, which is associated with asymptotic 
behavior of the function $K_3$ as  
$ 
K_3 \to (1-\cos \theta)/\sin \theta + \mu \sin\theta/r
$ \cite{KKS}. For a MA pair, for example, the integrated magnetic charge 
is zero, however the charge density distribution 
$g(x) = \frac{1}{2} {\rm Tr} \, \left( F_{ij} D_k \Phi \right)\varepsilon_{ijk}$
is not trivial, it has a maximum associated with node of the Higgs field 
on positive $z$-axis and symmetrically    
located minimum associated with the node on negative $z$-axis. The MA pair has 
non-vanishing magnetic dipole moment which can be relatively good evaluated by 
consideration of the magnetic 
charges as point charges located at the nodes \cite{KKS,S05}. 
On the other hand, the charge 2 monopole posses zero dipole moment since both nodes 
coincide.  

Therefore, we can conjecture that the MM pair possesses zero dipole moment. 
This condition implies that the boundary conditions on the functions, which enter the 
component $A_\varphi$ of the gauge potential, have to be modified and we impose  
\begin{equation}  \label{bound_MM}
K_1 \longrightarrow 0 \ , \ \ \ \
K_2 \longrightarrow - 1  \ , \ \ \ \
r^2 \partial_r K_3 \longrightarrow 0 \ , \ \ \ \ 
r^2 \partial_r K_4 \longrightarrow 0 \ , \ \ \ \ 
\end{equation}
We find that this modification yields new branch of gravitating MM solutions. 

Finally, regularity on the $z$-axis requires 
$ 
\partial_\theta f = \partial_\theta m =
   \partial_\theta l =0,  
$
whereas the matter field functions satisfy
$
K_1 = K_3 = \Phi_2 =0, \ 
\partial_\theta K_2 = \partial_\theta K_4 = \partial_\theta \Phi_1 =0  
$
for all these configurations.   

\section{Numerical results}
Subject to the above boundary conditions (\ref{bound_MM}), 
we solve the system of 9 coupled non-linear partial differential equation 
numerically in compact radial coordinate $x=r/(1+r) \in [0:1]$. The numerical 
calculations are based on the Newton-Raphson iterative procedure \cite{FIDI}.

For a small non-vanishing values of  $\lambda$ and $\alpha$ we
obtain the new static solution with two nodes of the Higgs field on the 
$z$-axis, which smoothly evolves as coupling constants begin to vary. 
Furthermore, there is a 
limiting branch of gravitating monopoles as $\lambda = 0$. 
These solutions    
are quite different from the known MA configurations (see Fig \ref{f-1}), 
in particular, the metric function $f$ does not possess a minimum at the origin. 

To confirm that these solutions can be interpreted as the monopole-monopole pair,
let us consider the magnetic charge density defined by integrand in (\ref{charge}).
For the configuration which we are analysing, the charge density remains 
positive everywhere. It has a shape of two tori those maxima form two rings   
in planes parallel to the $xy$-plane, intersecting the symmetry axis close to the 
nodes of the scalar field (see Fig \ref{f-2}, left). 
The energy density of the configuration 
possesses two maxima on the $z$-axis associated with nodes of the scalar field 
(Fig \ref{f-2}, right).

We can evaluate the charge  
by  straightforward substitution of the numerical 
solutions  into the definitions above, the calculation gives $g=2$. Thus, the 
solution represent gravitationally bounded pair of monopoles.  
In the limiting 
case $\alpha \to 0$ the solution  approaches flat space limit where the bound state 
ceases to exist and monopoles set to be free. Separation between the monopoles 
in this limit depends on the value of the  Higgs self-coupling $\lambda$, it is minimal  
for BPS monopoles: $d_{min} = 6.53$ in scaled units.
As  $\alpha$ increases the monopoles approach the flat space limit on a larger  
separation, e.g.,  $d_{min} = 7.31$ for  $\lambda - 0.1$.  
\begin{figure}\lbfig{f-1}\begin{center}
  \includegraphics[height=.35\textheight, angle =-90]{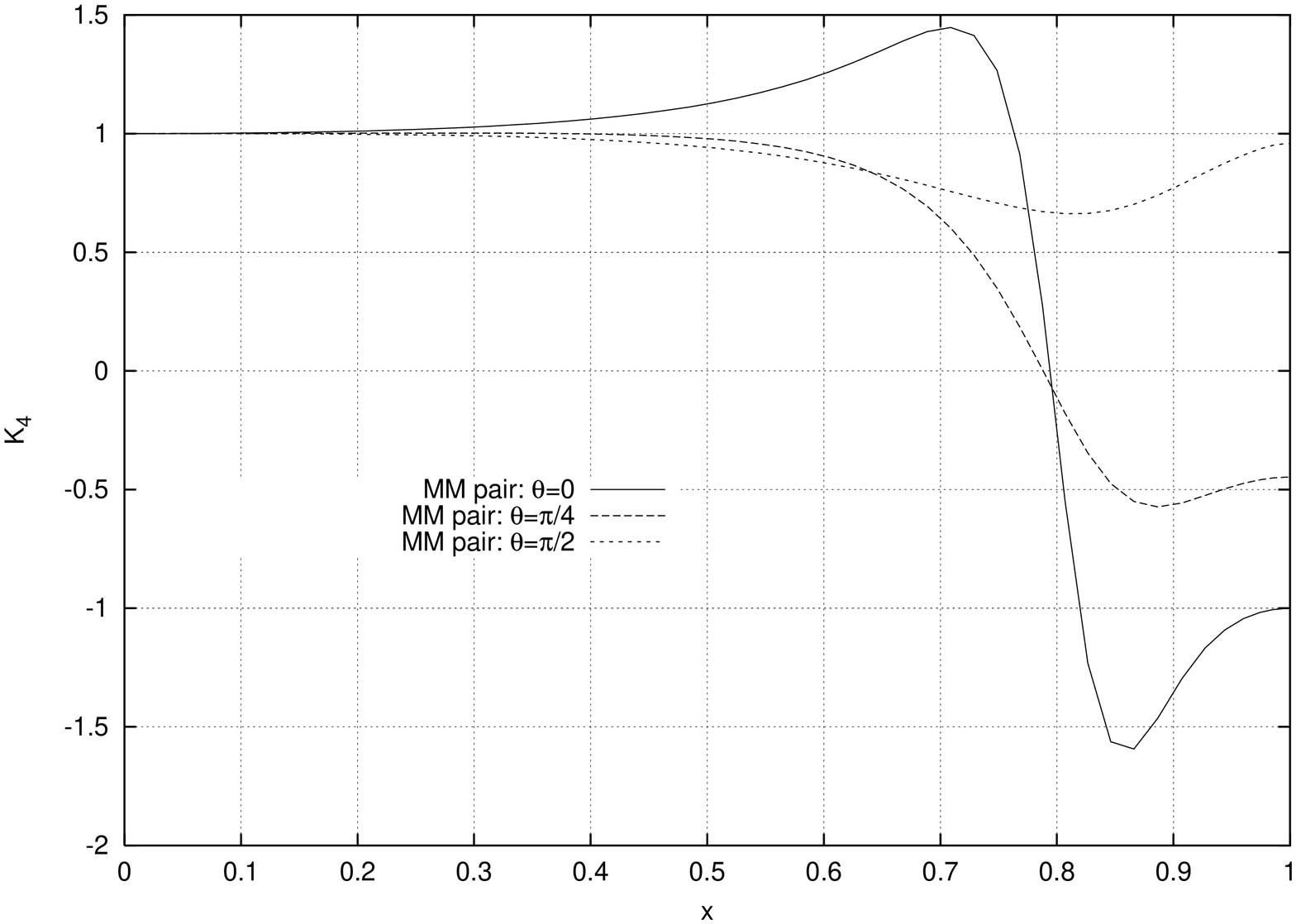}
\includegraphics[height=.35\textheight , angle =-90 ]{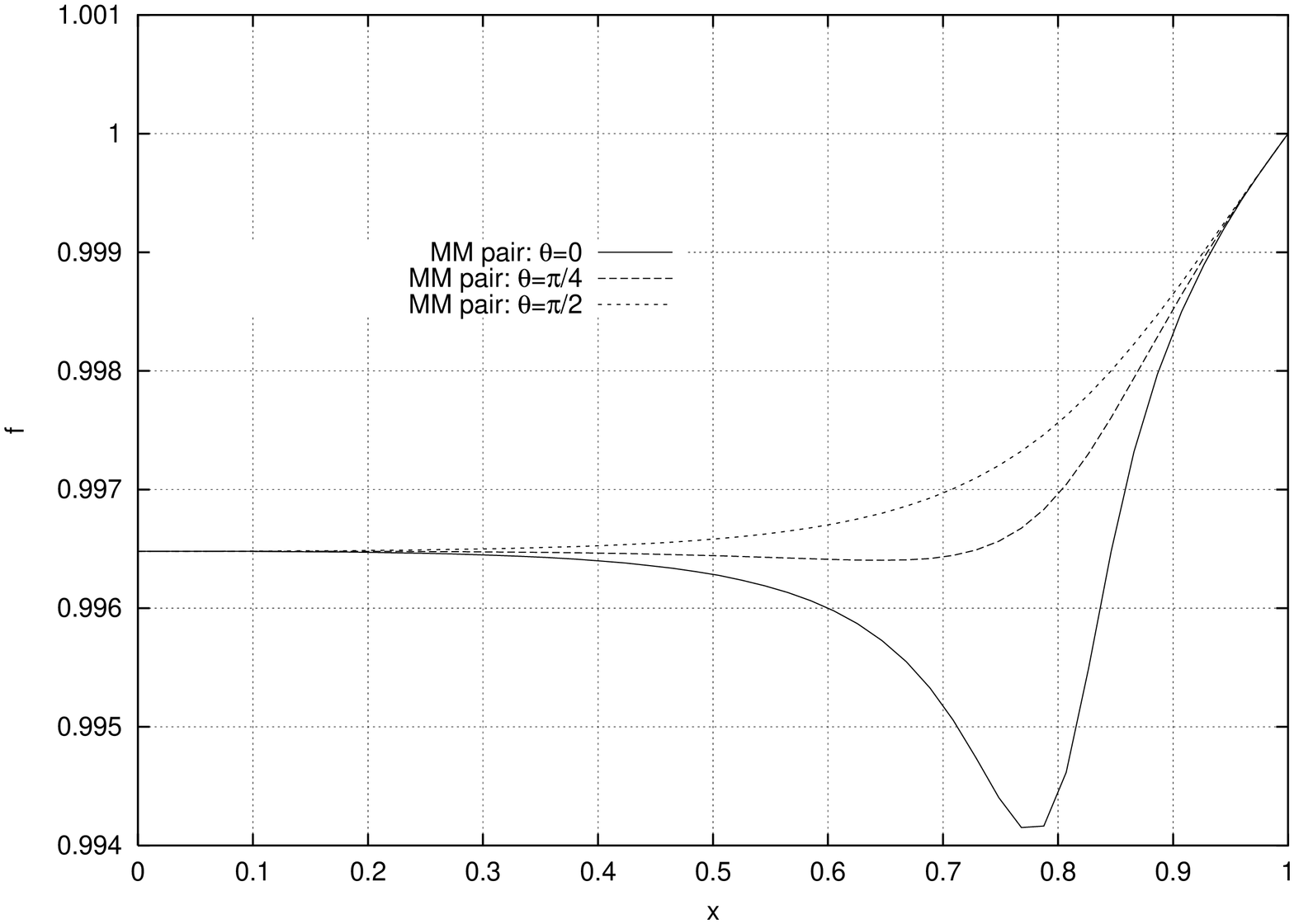}
\end{center}
  \caption{The gauge field function $K_4$ (left) and the metric function of the lower branch 
monopole-monopole pair solution are shown at $\lambda=0.5$, $\alpha = 0.05$
}
\end{figure}
 
\begin{figure}\lbfig{f-2}\begin{center}
  \includegraphics[height=.35\textheight, angle =-90]{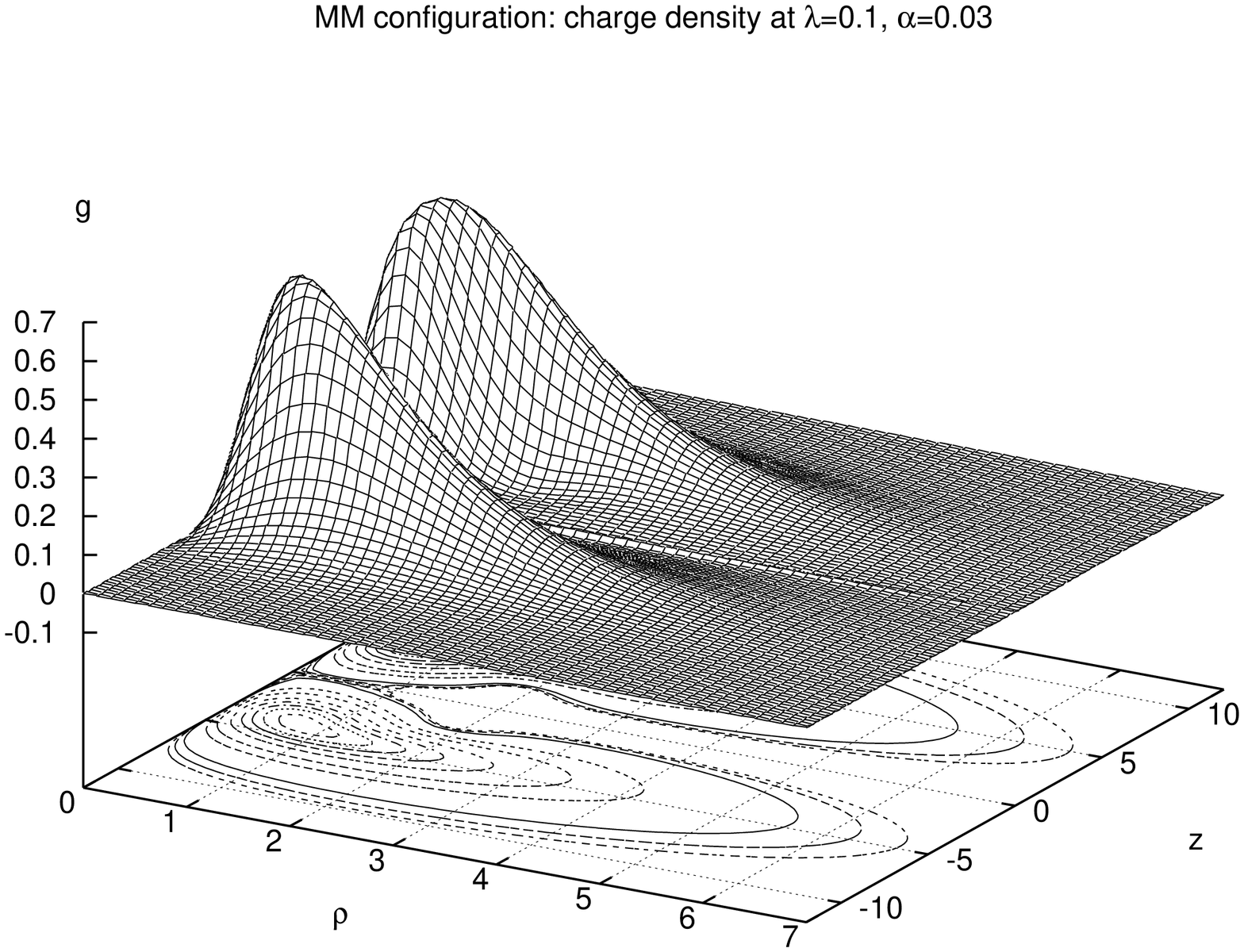}
  \includegraphics[height=.35\textheight, angle =-90]{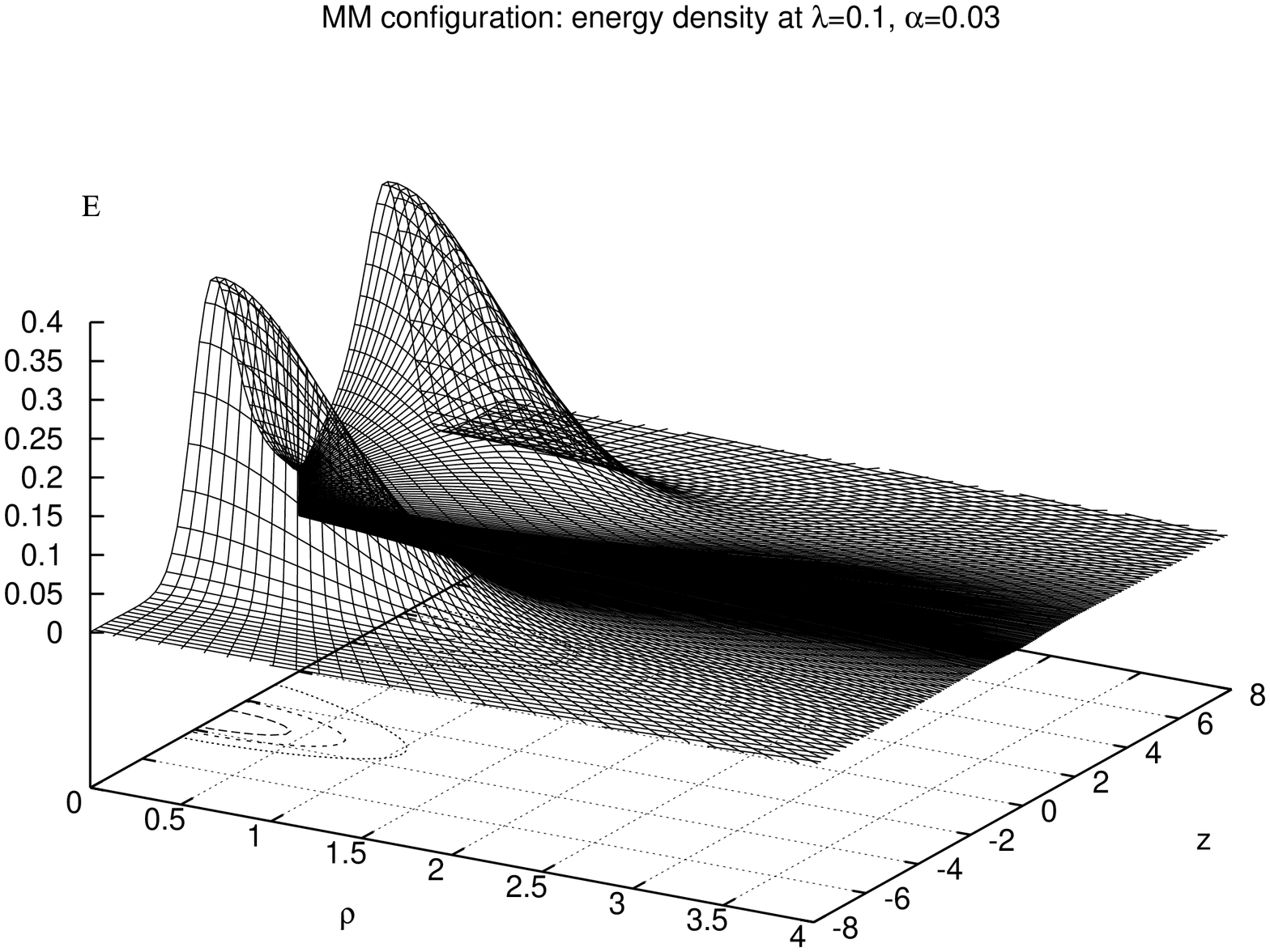}
\end{center}
  \caption{The charge density (left) and the energy density (right) distributions 
of the monopole-monopole pair solution 
are shown as functions of the coordinates $z$, $\rho = r \sin \theta$  
at $\lambda=0.1$, $\alpha = 0.03$. 
}
\end{figure}
This branch of gravitating monopoles extends up to a maximal value $\alpha_{cr}$, where 
it merges with a second branch similar to the case of the gravitating MA pair 
\cite{MAP}. As scalar coupling increases the $\alpha_{cr}$ decreases, e.g.,   
for $\lambda=0$ we have $\alpha_{cr}=0.838$ while for $\lambda=0.1$ 
$\alpha_{cr}=0.616$ and for $\lambda=0.5$ $\alpha_{cr}=0.562$. 
The second branch extends back to $\alpha=0$, as seen in Fig  \ref{f-3}.  
In this limit  the mass diverges and the configuration shrink to zero size. However, 
rescaling the coordinate $x \to \alpha x$ and  the scalar field $\Phi \to \Phi/\alpha $ 
leads to a limiting solution with finite size and finite scaled mass \cite{mapKK}. 
The scaled mass of the MM pair and the nodes of the Higgs field as functions of the 
coupling constant $\alpha$
are exhibited in Fig \ref{f-3}. Note that there is a principal 
difference between 
the metric function $f$ of the gravitating MM pair and that of the MA pair: the 
minimum of the former function is clearly 
associated with position of nodes of the scalar field while 
the minimum of the latter function remains at the origin \cite{mapKK}.

\begin{figure}\lbfig{f-3}\begin{center}
  \includegraphics[height=.35\textheight, angle =-90]{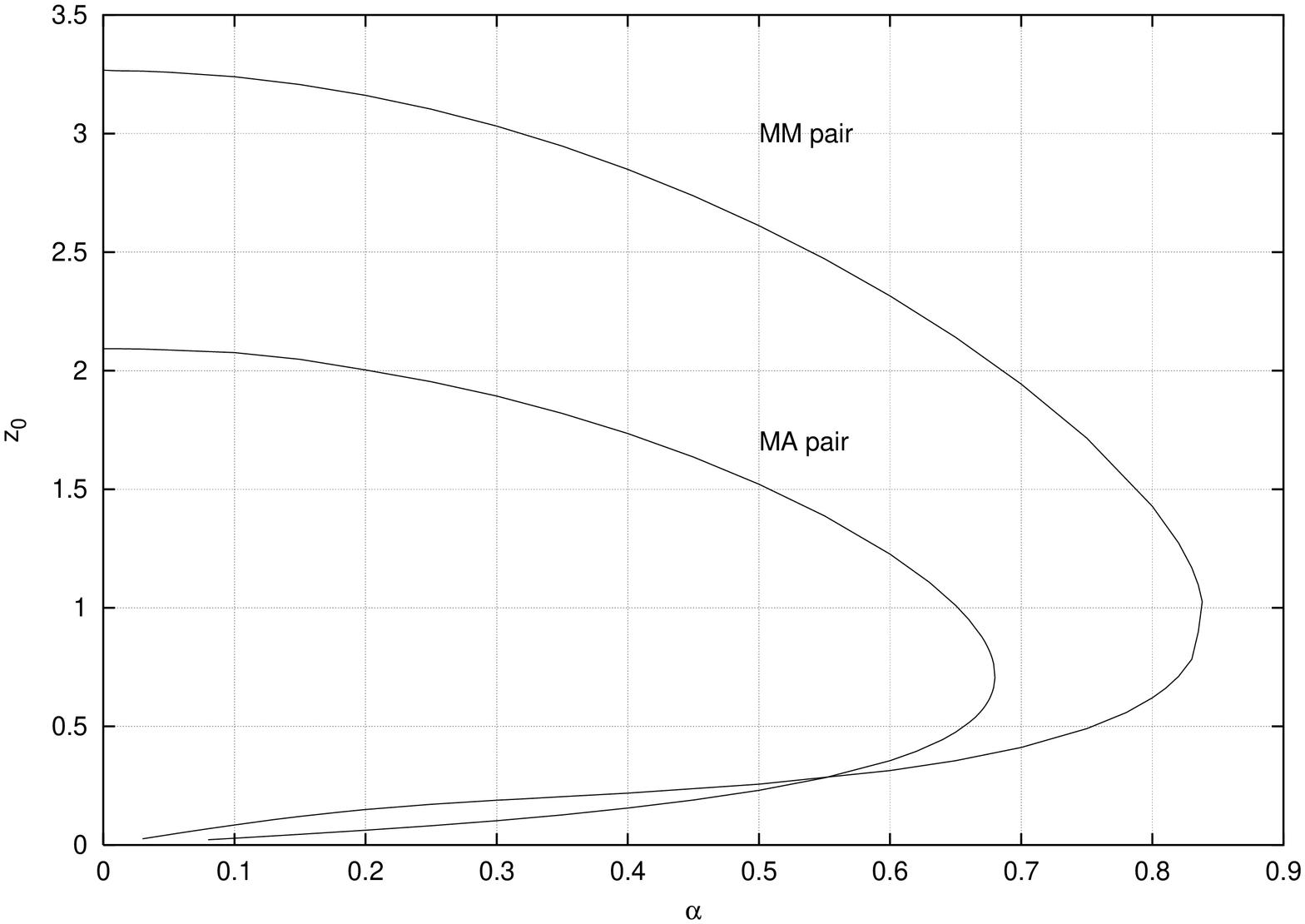}
  \includegraphics[height=.35\textheight, angle =-90]{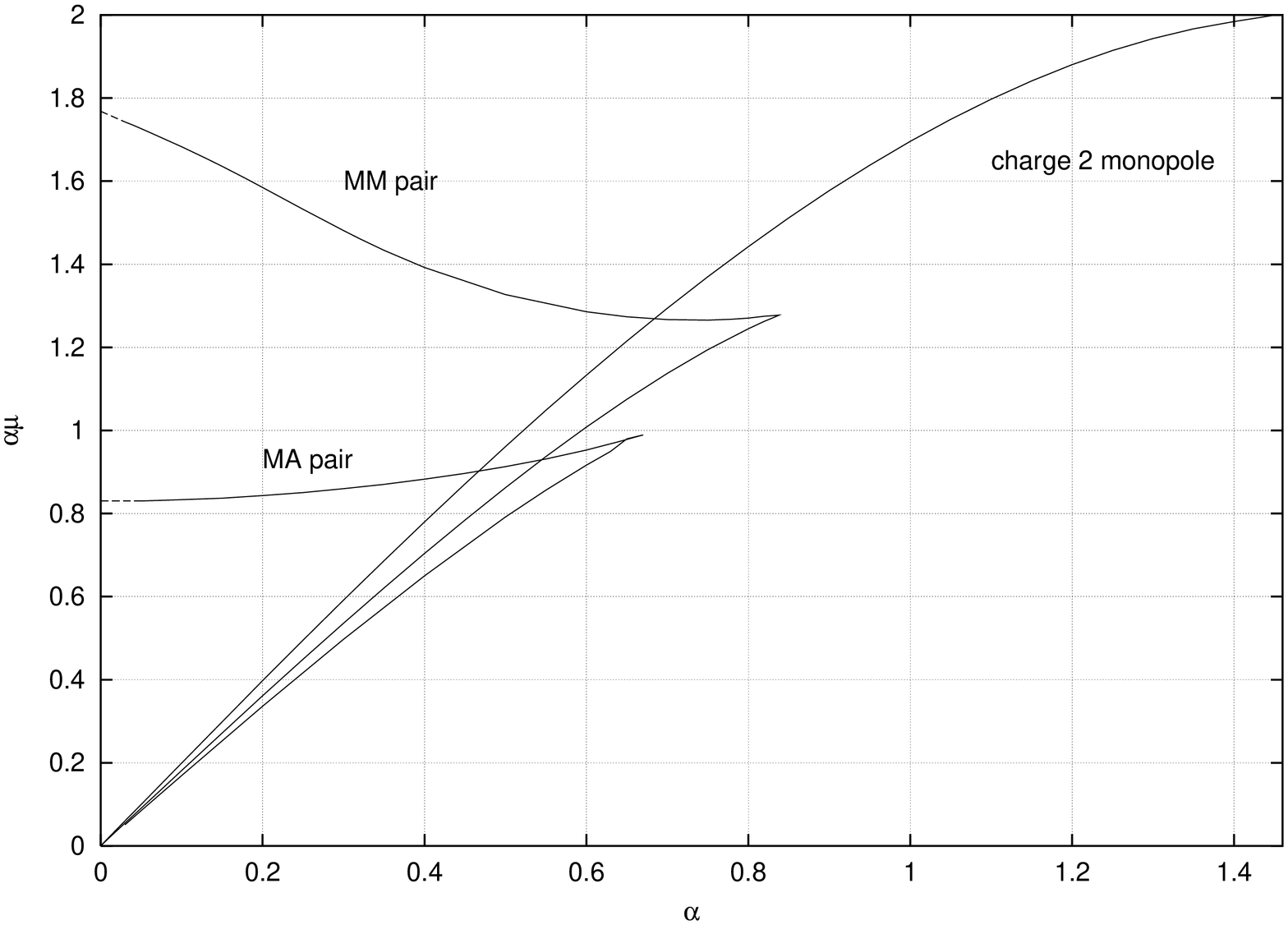}
\end{center}
  \caption{The nodes of the Higgs field are shown as 
functions of the coupling constant $\alpha$ for the MM pair and the MA pair solutions
at $\lambda = 0$  (left). Also the scaled masses of the MM pair, the MA pair and the 
charge 2 monopole solution are shown  as functions of the coupling constant $\alpha$ 
at $\lambda = 0$. 
}
\end{figure}
 
Considering the limit $\alpha \to 0$ on the upper branch in scaled coordinates, 
we observe that the solution may be thought of as composed of a scaled charge 2 
monopole solution in the inner region and the second Bartnik-McKinnon 
solution with two nodes of the gauge field function in the outer region, 
as seen in Fig  \ref{f-4}. The inner region shrinks to zero size as  $\alpha \to 0$. 

\begin{figure}\lbfig{f-4}\begin{center}
  \includegraphics[height=.35\textheight, angle =-90]{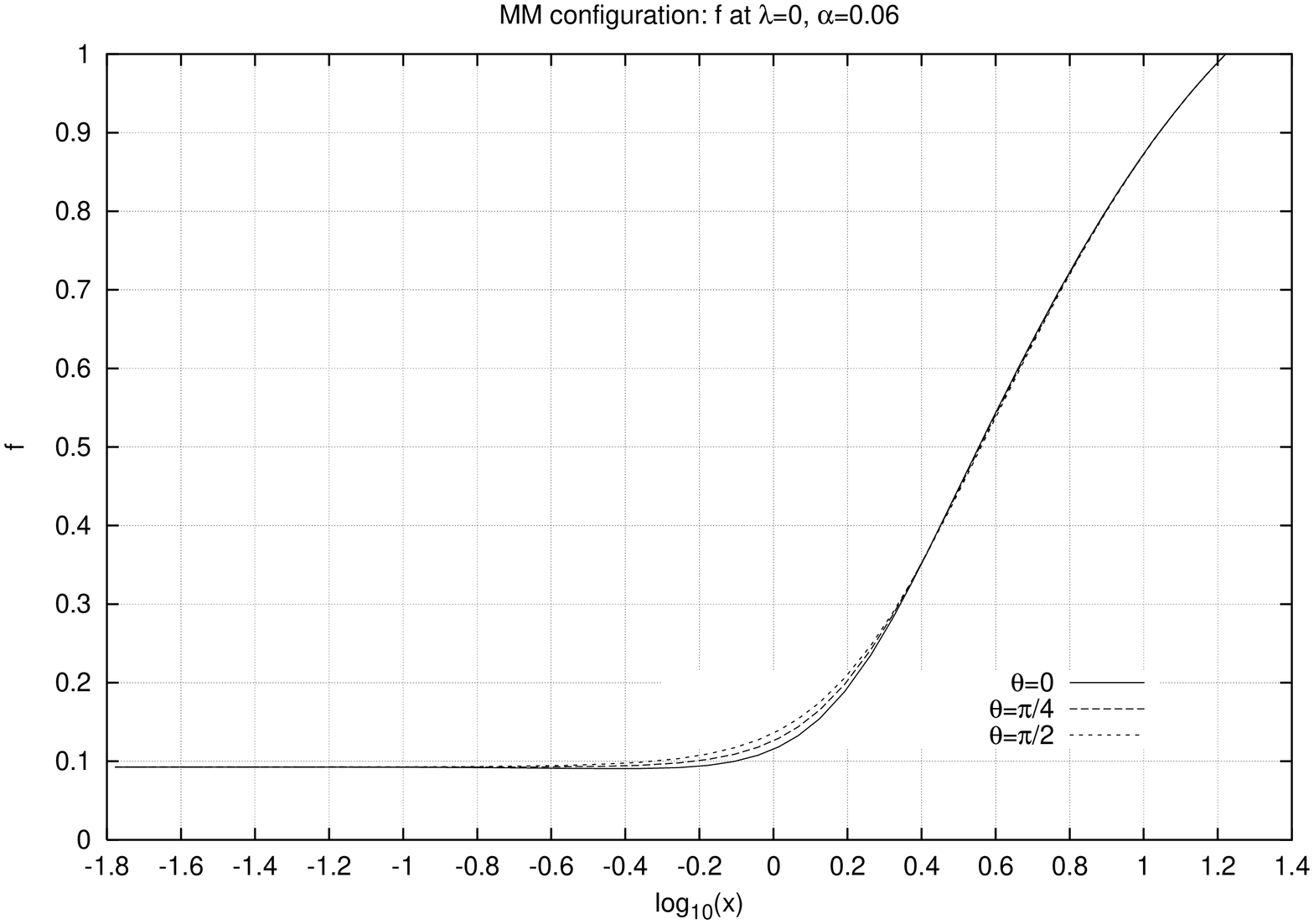}
  \includegraphics[height=.35\textheight, angle =-90]{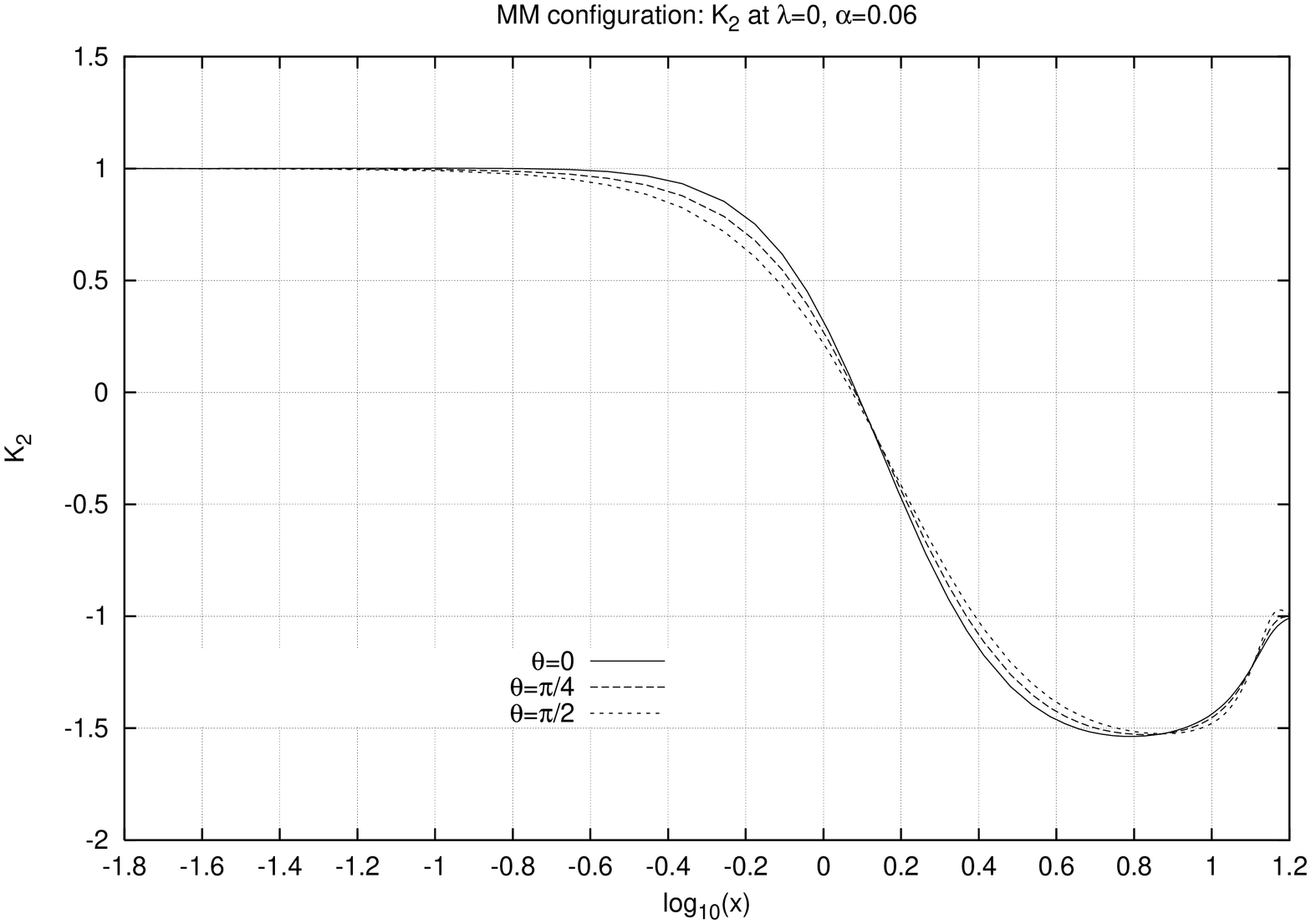}
\end{center}
  \caption{The metric function $f$ (left) and the gauge field function $K_2$ (right) 
of the MM pair solution are shown in scaled coordinates at  $\alpha = 0.06$ and 
$\lambda = 0$. 
}
\end{figure}

Comparing the solutions with the gravitating charge 2 monopole \cite{HKK} we observe 
that on the lower mass branch, the mass of the monopole-monopole pair for the same 
value of $\lambda = 0$ 
is lower (see Fig \ref{f-3}). Since both configurations are in the same topological 
sector with charge 2, one may expect that gravitating 2-monopole is unstable. 
Hovewer, these solutions evolve differently, as $\alpha$ increases, 
the branch of gravitating 2-monopole
approaches the extremal Reissner-Nordstr\"om black hole with magnetic 
charge 2 \cite{gmono}
while both gravitating MM pair and gravitating MA pair evolve back on the upper 
branch and they are linked to the 
Bartnik-McKinnon solutions of pure EYM theory. 

In is instructive to compare the limiting behavior of these three 
different configurations, all possessing  two nodes of the 
Higgs field. In Fig \ref{f-5} we exhibit  the corersponding gauge field functions $K_2$ 
and the metric functions $f$ for the upper branch MM pair solution, the upper branch 
MA pair solution and charge 2 monopole.      

\begin{figure}\lbfig{f-5}\begin{center}
  \includegraphics[height=.35\textheight, angle =-90]{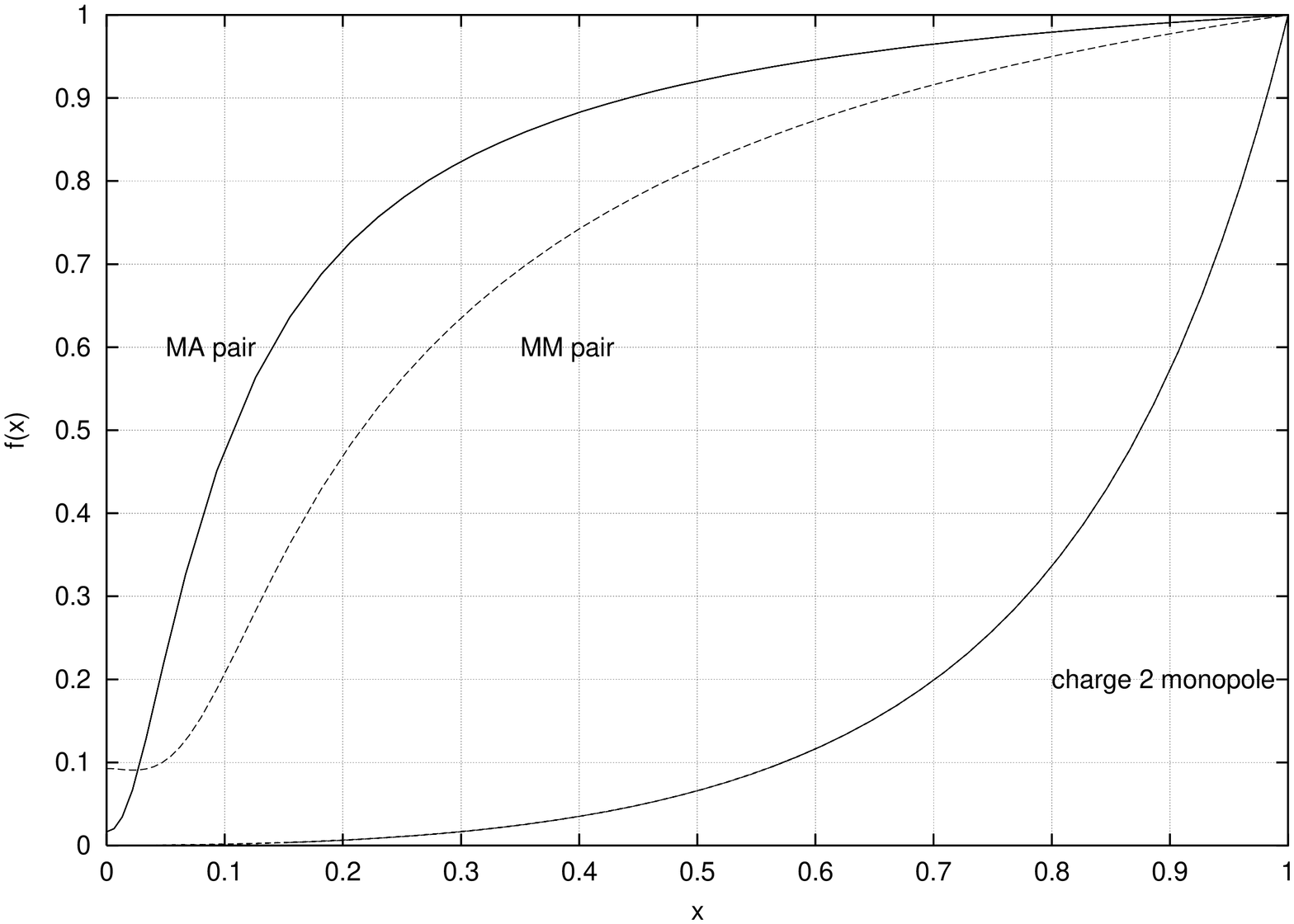}
  \includegraphics[height=.35\textheight, angle =-90]{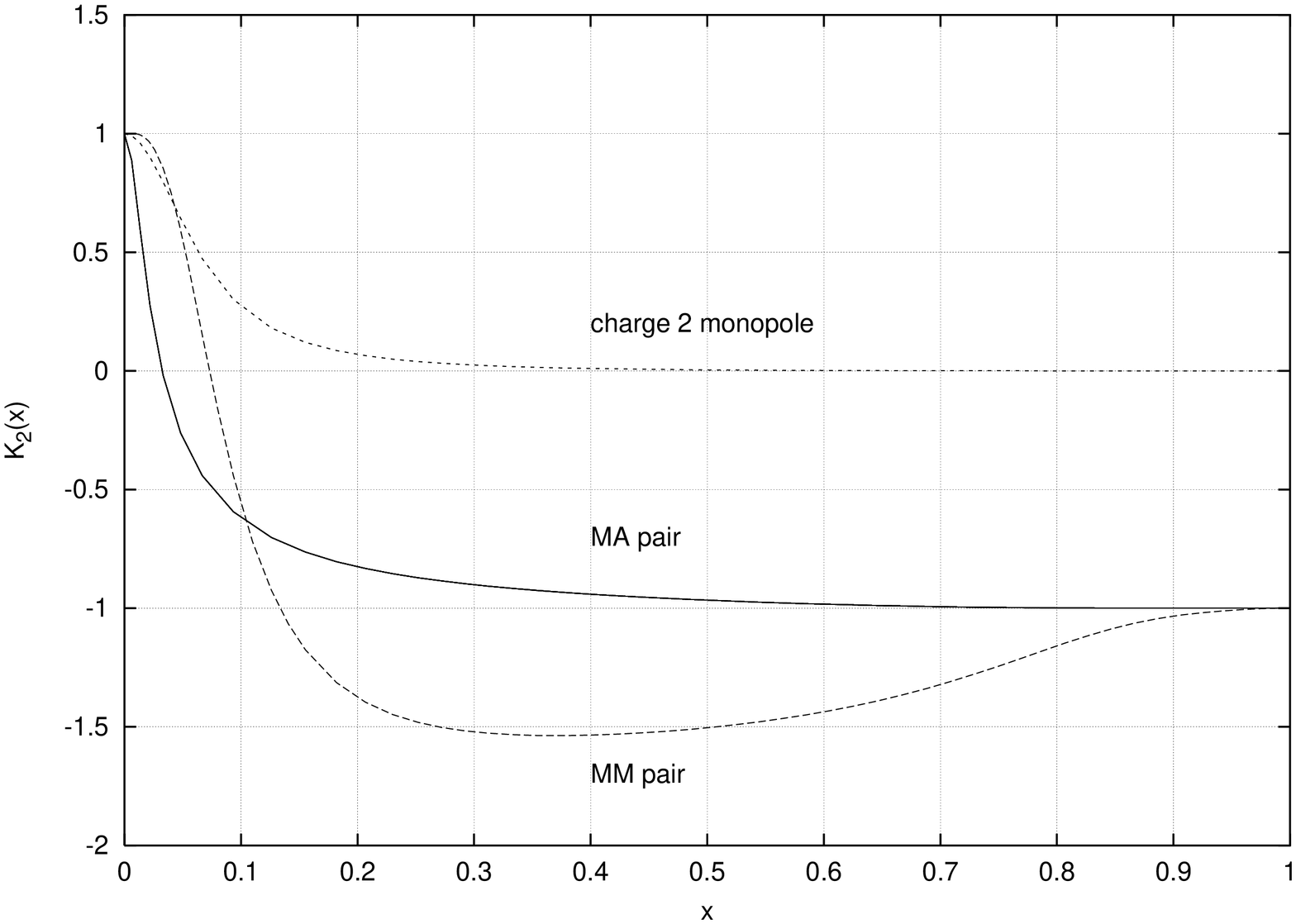}
\end{center}
  \caption{The metric function $f$ (left) and the gauge field function $K_2$ (right) 
of the upper branch MM pair solution, the upper branch 
MA pair solution and the charge 2 monopole are shown at $\lambda = 0$ and  
$\alpha = 0.05$ (MM and MA configuration) 
and at  $\alpha = 1.45$ (charge 2 monopole)
}
\end{figure}

Concluding, we have found new static axially symmetric solutions of
$SU(2)$ EYMH theory, which represent gravitating monopole-monopole pairs. 
This results holds both for zero and for finite Higgs self-coupling \cite{long}.
We expect that there are also gravitating 
multiply magnetically charged solutions which are counterparts 
of the vortex rings discussed in \cite{KKS}. Since we observe that for the 
MM pair the minimum of the metric function $f$ coincides with location of the monopoles,
some interesting features of the black hole solutions related to these configuration 
may be observed.  Study of these and other solutions of the new class, 
only simplest of those was discussed in this note, 
will be presented elsewhere \cite{long}.

\end{document}